\documentclass[11pt]{asaproc}

\usepackage{times}
\usepackage{amsmath}
\usepackage[pdftex]{graphicx}
\usepackage{psfrag,epsf}
\usepackage{enumerate}
\usepackage[round]{natbib}   
\usepackage{url} 
\usepackage{algorithm,algpseudocode}
\usepackage{lscape}
\usepackage[stable]{footmisc}
\usepackage{parskip}
\usepackage{xfrac}
\usepackage{subcaption}
\usepackage{alphalph}

\renewcommand{\thefigure}{\arabic{figure}}

\title{Correcting for Reporting Delays in Cyber Incidents}

\author{Seema Sangari\thanks{School of Data Science and Analytics, Kennesaw State University, 3391 Town Point Dr. NW,Kennesaw, GA 30144. Contact: SSangar1@students.kennesaw.edu} \and  Dr. Eric Dallal\thanks{AIR Worldwide, Verisk Cyber Solutions, Lafayette City Center,
		2 Avenue de Lafayette, 2nd Floor, Boston, MA 02111. Contact: Eric.Dallal@verisk.com}}

\begin{document}
	
	\maketitle
	
	\begin{abstract}
		With an ever evolving cyber domain, delays in reporting incidents are a well-known problem in the cyber insurance industry. Addressing this problem is a requisite to obtaining the true picture of cyber incident rates and to model it appropriately. The proposed algorithm addresses this problem by creating a model of the distribution of reporting delays and using the model to correct reported incident counts to account for the expected proportion of incidents that have occurred but not yet been reported. In particular, this correction shows an increase in the number of cyber events in recent months rather than the decline suggested by reported counts.
	\end{abstract}
	
	\section{Introduction}
	\label{sec:intro}
	The cyber security domain is evolving rapidly, with new attack vectors emerging regularly, and even the most up-to-date data cannot be considered complete. Cyber incidents take a long time to become known and even longer to appear in online databases. While some cyber events are known immediately after they occur, most events are often reported many months or years after the event actually occurred, resulting in biased data. Marriott’s major cyber incident occurred in 2014 but was only reported in 2018. Major cyber events become headlines in leading newspapers when reported publicly rather than when they happened. Smaller cyber events may never be reported at all, or have extreme delays, as only public companies and those with personally identifiable information may be obligated to report. Sometimes reporting can take 5-10 years, for various intentional or unintentional reasons - failing to realize that a cyber incident happened, failing to immediately determine the extent of accessed or stolen data, or deciding not to publicize the incident for fear of reputation risk and consequent financial impacts. As a result, reporting delays are often observed in historical cyber event databases. These databases show a decrease in cyber incidents in the recent few years. 
	 \cite{DerryckColeman2021} raised the concern that cyber incidents would remain undetected due to advanced threat techniques.
	
	Cyber risk modeling firms rely upon historical data to build their models, which are in turn relied upon by cyber insurers for underwriting, portfolio management, and risk transfer. To build robust loss estimation models for today's evolving cyber world, the most recent and updated information is required, with as little bias as possible. Correcting reporting delays in these databases is therefore a key requirement to have trustworthy cyber insurance models. With the necessary corrections, one can then properly examine temporal trends in the targeting of industries or in attacker tactics.
	
	Terminology followed in this paper:
	\newline
	$i$: Incident \newline
	$I$: Set of all incidents  \newline
	$delay, \delta_i $: Time between the incident date and the reporting date  \newline
	$age, A_i$: Time between the incident date and the last incident reporting date in the data  \newline
	$f_\Delta$: Probability density function of the $delay$ distribution  \newline
	$F_\Delta$: Cumulative distribution function of the $delay$ distribution  \newline	
	\section{Literature Review}
	\label{LitRev}
	Most of the literature on reporting delays addresses the medical space. We are not aware of any existing papers that propose a methodology for correcting cyber incident counts to account for reporting delays. However, \cite{DerryckColeman2021} does examine both the distribution of the number of days to discover cyber incidents and the number of days to disclose them.
	
	\cite{Harris} described reporting delays as a statistical problem for the first time. \cite{Heisterkamp1988a, Heisterkamp1988, Heisterkamp1989} and \cite{ Brookmeyer1989} made distributional assumptions and built linear/quadratic models whereas \cite{Rosenberg1990} and \cite{Cheng1991} suggested Poisson models. These are easy to implement but assume stationary reporting delays and don't capture trends. \cite{Morgan1986}, \cite{Downs1987, Dow1988}), \cite{Healy1988} and   \cite{Heisterkamp1988a, Heisterkamp1988, Heisterkamp1989} fitted  exponential, integrated logistic and log-linear models to capture trends. \cite{Gail1988}, \cite{Brookmeyer1990}, \cite{Kalbfleisch1991} and \cite{Esbjerg1999} applied conditional probabilities to capture trends but this resulted in over-fitting. \cite{Lawless1994} proposed a multinomial model with distributed random effects based on Dirichlet/Poisson/Gamma distributions to capture trends in a timely fashion but failed to handle longer delays.
	
	
	\cite{Wang1992} suggested maximum likelihood estimation (MLE) based non-parametric and semi-parametric approaches but with complete\footnote{Complete data - No further events are expected to be reported with delays.} data. \cite{Harris2020} suggested correcting COVID cases with an expectation-maximization (EM) algorithm and trained the model with complete data to correct test data. \cite{White2009} and \cite{Weinberger2020} proposed a simpler method based on proportions but also require complete data to train.
	
	\cite{Wang1992}, \cite{Keiding1992} and \cite{Midthune2005} applied truncated models to avoid random effects but require stable reporting delays.
	
	\cite{Hohle2014}, \cite{Bastos2019} and \cite{Chitwood2020} suggested a Bayesian and hierarchical approach with Poisson and negative binomial distributions. It is easy to implement but makes distributional assumptions.
	\cite{Noufaily2015, Noufaily2016}) suggested a log-likelihood approach with a truncation model that is data driven but sensitive to the choice of three reporting time steps.
	
	\cite{Bastos2019} suggested a chain-ladder approach but it is sensitive to outliers.
	
	\cite{Jewell1989}, \cite{Zhao2009}, \cite{Zhao2010}, and \cite{Avanzi2016} 
	investigated cyber claims data to account for reporting delays from a capital reserving perspective. This problem is different from the one being investigated, since reporting delays in claims are due only to detection delays.
	
	As stated in \cite{Brookmeyer1990}, none of these approaches deal with delays longer than any previously observed.
	
	\section{Data}
	\label{sec:Data}
	The data used is a proprietary data set constructed by merging multiple source cyber event sets together. The source cyber event sets contained differing fields that, at minimum, provided information as to the ``what'', ``when'' and ``who'' of an event. Concretely, the data sets included a description of the event, the name of the company to which the cyber event occurred (N.B.: aggregation events separately list each company known to be impacted), along with the date that the event occurred and the date that the event was reported.
	
	Because company names frequently differ from one data set to another, the data sets could not be de-duplicated based on direct string matching. Instead, the event data sets were matched to a firmographic data set containing approximately 55 million businesses in the US. This was done via a previously developed matching algorithm that examined company name, industry classification (e.g., via NAICS codes), address information, and any other fields common to both the cyber event data set and the firmographic data set. Events in distinct cyber event data sets were identified as the same when:
	\begin{enumerate}
		\vspace{0cm}
		\setlength\itemsep{-0.5em}
		\item They were matched to the same company in the firmographic data set; and
		\item They were listed as having occurred within 1 week of each other.
	\end{enumerate}
	
\begin{figure}[htbp]
		\centering
		\begin{subfigure}{.5\textwidth}
			\centering
			\fbox{\includegraphics[scale=0.295, angle=90,origin= c]{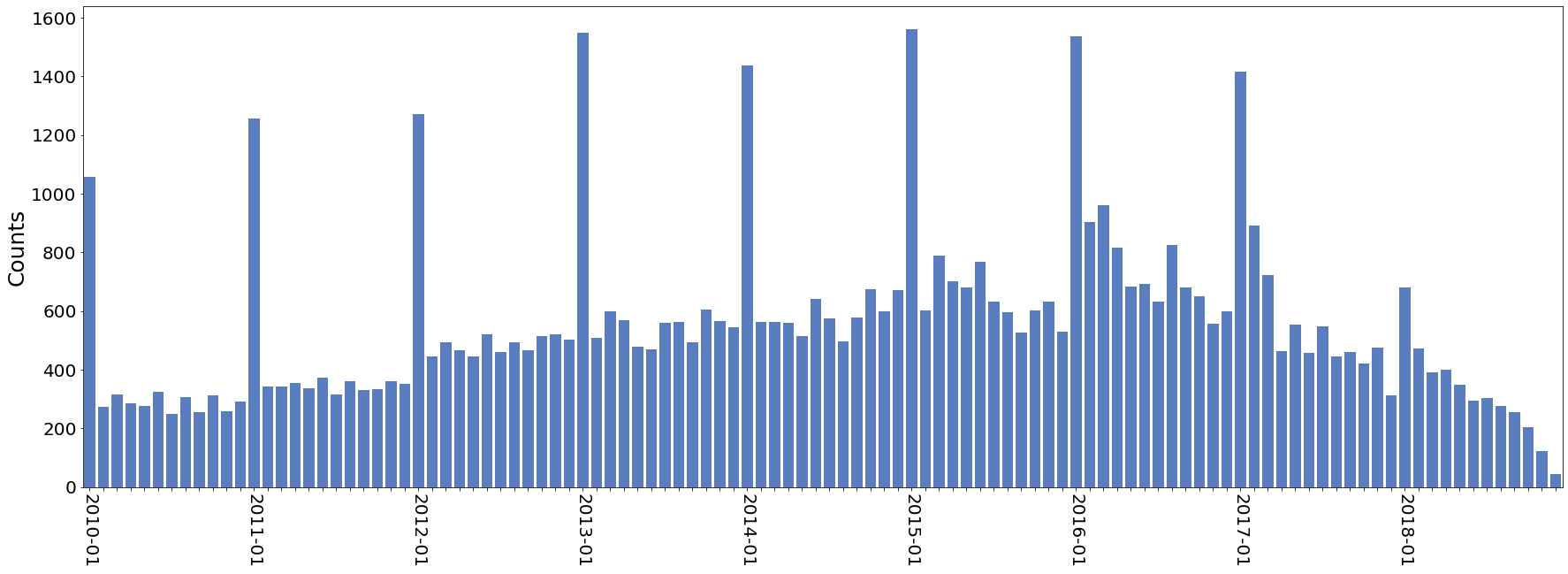}}
			\caption{With Default Date}
			\label{fig:DefaultDateCounts}
		\end{subfigure}%
		\begin{subfigure}{.5\textwidth}
			\centering
			\fbox{\includegraphics[scale=0.295,  angle=90,origin= c]{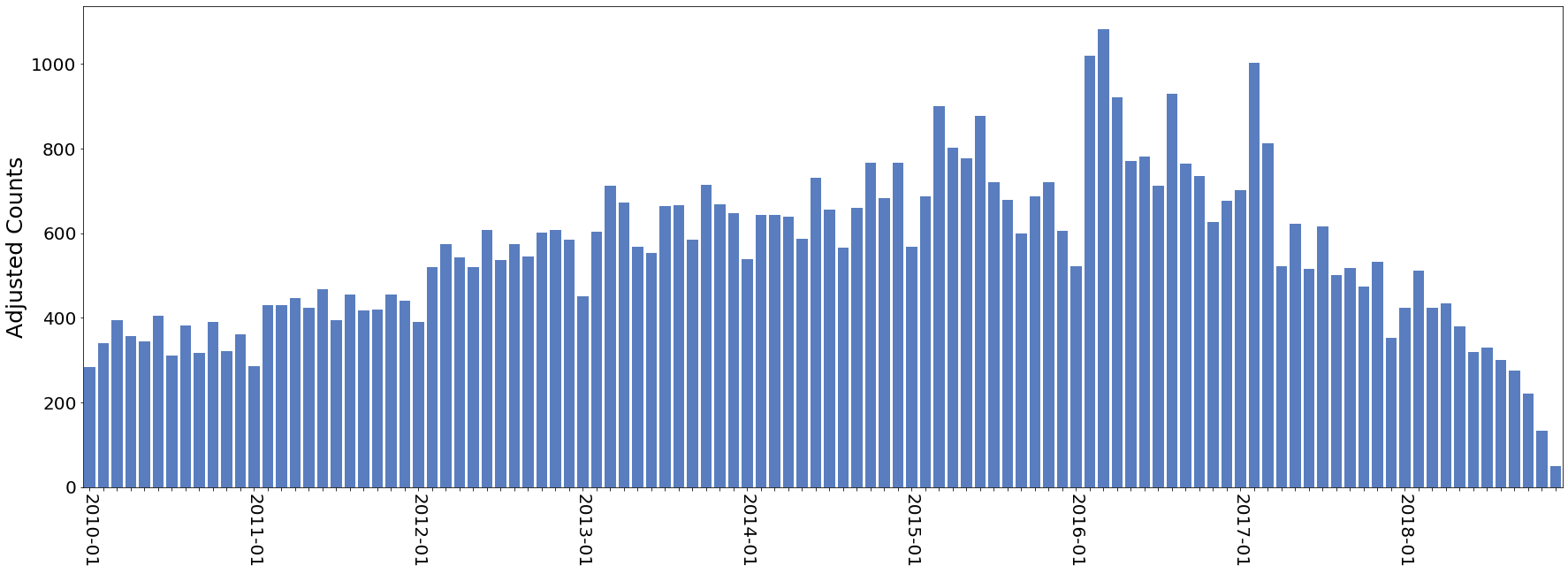}} 
			\caption{Adjusted for Default Date}
			\label{fig:AdjustedDefaultCounts}
		\end{subfigure}
		\vspace{-15pt}
		\caption{Cyber Event Counts until December 2018}
		\label{fig:OriginalCounts}
	\end{figure}

	\textbf{Limitations}: Some events do not have an occurrence date listed, and were therefore excluded from this analysis. There are also large spikes in event counts listed as having occurred on January 1st in most years (Fig.~\ref{fig:DefaultDateCounts}), which was assumed to be a default value when only the year of the event was known. These events were therefore re-distributed proportionally throughout the year as shown in Fig.~\ref{fig:AdjustedDefaultCounts} but excluded while developing the approach.
	
	\section{Proposed Approach}
	\label{sec:Approach}
	The proposed approach consists of estimating the reporting delay distribution $f_{\Delta}$ from the empirical data. With this distribution, a corrected count of events with age $a$ can be determined by dividing the raw counts by $F_{\Delta}(a)$, as the latter expression represents the proportion of events that are reported within a delay, $\delta$, of less than $a$ or, equivalently, the proportion of events that are reported as of the present. There are four complications with estimating the delay distribution in the ``obvious'' way:
	
	\begin{itemize}
		\item{The nature of reporting delays means that direct estimates from empirical data will be biased towards shorter delays, since recent events could only appear in the data set in the first place if the reporting delay is small.}
		\item{The estimates assume zero probability of any delay longer than the longest in the data set.}
		\item{The estimates for long reporting delays are based on few data points.}
		\item{The reporting delay distribution may not be stationary.}
	\end{itemize}
	
\begin{figure}[H]
		\centering
		\begin{subfigure}{.5\textwidth}
			\centering
			\fbox{\includegraphics[width=0.95\linewidth]{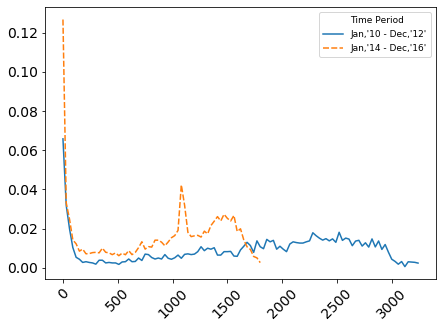}}
			\caption{Probability Distribution}
			\label{fig:sub1}
		\end{subfigure}%
		\begin{subfigure}{.5\textwidth}
			\centering
			\fbox{\includegraphics[width=0.95\linewidth]{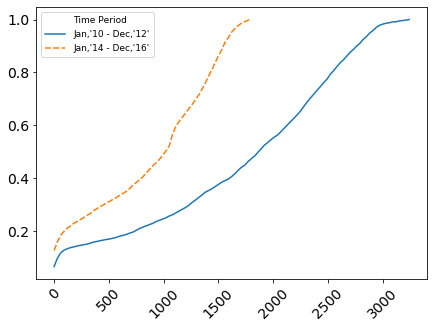}}
			\caption{Cumulative Distribution}
			\label{fig:sub2}
		\end{subfigure}
		\vspace{-10pt}
		\caption{PDF and CDF of  Delay Distribution generated for Dec.,'12 and Dec.,'16}
		\label{fig:NonStationarity}
	\end{figure}

	Mathematically, the first problem is that, for incidents with age $a$, the delay $\delta_i$ can only be observed if $\delta_i <= a$. This problem can be resolved if one assumes that $f_{\Delta}$ is stationary, in which case the distribution can be estimated for delays, $\delta$, larger than $a$ from older events. This is the approach taken in Algorithm~\ref{alg:ComputeDelayDistribution}. To test for stationary, the algorithm was run on two year windows of the data set centered at different times. (Note that the two year windows include all events that \emph{occur} within the window, irrespective of when they are reported.) This showed that the delay distribution is non-stationary, as can be seen in Fig.~\ref{fig:NonStationarity}.
	
	To deal with this non-stationarity, the delay distribution was modeled via parametric distributions where parameters were estimated over monthly rolling two year windows. For each two year window, optimal parameters were determined based on matching cumulative distributions. Specifically, Algorithm~\ref{alg:ComputeDelayDistribution} was run on each of the monthly two year windows to obtain an estimate of the delay distribution for that time window. For a given two year window and modeled distribution, an optimization algorithm was used to determine the parameters that gave the best match between:
	\begin{itemize}
		\item{the delay distribution obtained by Algorithm~\ref{alg:ComputeDelayDistribution}; and}
		\item{the parametric distribution restricted to the domain where the maximum delay is less than the window's maximum reporting delay.}
	\end{itemize}
	
	The optimal distribution parameters for each window were computed using a derivative free optimizer.
	
	In what follows, the above two distributions will be referred to as the ``debiased empirical delay distribution'' and the ``modeled delay distribution'', respectively.
	
	\subsection{Generating the Debiased Empirical Delay Distribution} \label{ssec: DelayDist}
	Inspired from \cite{Brookmeyer1990}, the proposed algorithm works on the limitation that the delay, $\delta$, is not considered beyond age and hence only a distribution conditioned on the delay being less than or equal to age can be estimated. The algorithm corrects the incident counts with $F_\Delta$ estimated from empirical data. The algorithm applies a top down approach to estimate the distribution ``from the outside in'', accounting for the estimated proportion of events that have occurred but not yet been reported as the distribution itself is being computed.
	
	Let $A_{\max} := \max_{i \in \mathcal{I}}A_i$ be the maximal age of any event in the data set. Also let $h_A(a)$ be the number of incidents of age $a$, and let $h_{\Delta}(\delta)$ be the number of incidents with delay $\delta$. Formally,
	\begin{eqnarray}
		h_A(a) & = & |\{i \in \mathcal{I} : A_i = a\}| \label{eq:ha} \\
		h_{\Delta}(\delta) & = & |\{i \in \mathcal{I} : \Delta_i = \delta\}| \label{eq:hdelta}
	\end{eqnarray}
	
	Then the equation to estimate the delay distribution is:
	\begin{align}
		f_{\Delta}(\delta) = \frac{h_{\Delta}(\delta)}{\sum_{a = \delta}^{A_{\max}}\sfrac{h_A(a)}{F_{\Delta}(a)}} \label{eq:delay_dist}
	\end{align}
	
	Intuitively, the distribution is generated based on the ratio of the number of events with the given delay period,  $h_{\Delta}(\delta)$, to the best estimate of the true number of events whose age is old enough to be seen in the incident data set (i.e., where the age is larger than the delay under consideration), $\sum_{a = \delta}^{A_{\max}}\sfrac{h_A(a)}{F_{\Delta}(a)}$. Algorithm~\ref{alg:ComputeDelayDistribution} shows the algorithmic implementation to generate the distribution.
	
	\begin{algorithm}[H]
		\begin{algorithmic}[1]
			\Require{The histograms, $h_{A}$ and $h_{\Delta}$, computed as in Eqs.~(\ref{eq:ha}) and (\ref{eq:hdelta}), respectively.}
			\Ensure{The distribution $f_{\Delta}$.}
			\Statex
			\Function{ComputeDelayDistribution}{$h_{A}$, $h_{\Delta}$}
			\State{$A_{\max} \gets \max_{i \in \mathcal{I}}A_i$}
			\State{$F_{\Delta}(A_{\max}) \gets 1$}
			\State{$\delta_{max} \gets \max_{i \in \mathcal{I}}\delta_i$}
			\For{$\delta \gets A_{\max}$ to $\delta = 0$}
			\State{$den \gets 0$}
			\For{$a \gets \delta$ to $\delta_{\max}$}
			\State $den \gets den + h_{A}(a)/F_{\Delta}(a)$ \Comment{Computes denominator}
			\EndFor
			\State{$f_{\Delta}(\delta) \gets h_{\Delta}(\delta)/den$} \Comment{Computes PDF}
			\State{$F_{\Delta}(\delta-1) \gets F_{\Delta}(\delta) - f_{\Delta}(\delta)$}
			\Comment{Updates CDF}
			\State{$\delta_{max} \gets \delta$}
			\EndFor
			\State \Return $f_{\Delta}$
			\EndFunction
		\end{algorithmic}
		\caption{Algorithm for computing the debiased empirical delay distribution}
		\label{alg:ComputeDelayDistribution}
	\end{algorithm}
	\vskip 1em
	
	\subsection{Generating the Modeled Delay Distribution}
	As previously described, the modeled delay distribution is determined by an optimization that minimizes the difference in the cumulative distribution functions of the debiased empirical delay distribution and the modeled delay distribution, restricted to the domain $[0,\delta_{\max}]$ (N.B.: $\delta_{\max}$ is the maximum observed delay). The PDF and CDF plots shown in Fig.~\ref{fig:NonStationarity} suggest that a single distribution will not provide a good fit, as the delay distribution is bi-modal. Rather, a mixture of distributions would be required. Fig.~\ref{fig:NonStationarity} suggests a mix of exponential and normal distributions.
	
	Mathematically, 
	\begin{align}
		F_\theta(\delta)&= \alpha F_{Exp}(\delta, Scale) + (1-\alpha) F_{N}(\delta, \mu, \sigma) \label{Eq: MixCDF}\\
		&\text{where     } F_{Exp} = \text{Exponential CDF with parameter, $Scale = \sfrac{1}{\lambda}$} \nonumber\\
		&\quad \quad \quad F_N = \text{Normal CDF with parameters, $\mu$ and  $\sigma$}  \nonumber
	\end{align}
	
	A possible interpretation of the bi-modal nature of the reporting delay distribution is that there are two distributions: one for events that are discovered almost immediately (modeled by the exponential) and one for events which have a delay due to both discovery time and public disclosure time (modeled by the normal). The parameter $\alpha$ can therefore be interpreted as the proportion of events that are discovered right away by the organization.
	
	Since the normal distribution is defined on $(-\infty, \infty)$ and reporting delays cannot be negative, the modeled delay distribution needs to be adjusted. The CDF in Eq.~\ref{Eq: MixCDF} truncated to $[0,\infty)$ (and re-normalized) can be expressed as
	\begin{align}
		F_\theta(\delta)=&\frac{\alpha (F_{Exp}(\delta, Scale)) + (1-\alpha)\overbrace{(F_{N}(\delta, \mu, \sigma)-F_{N}(0, \mu, \sigma)}^{\text{Truncated Normal Distribution until $\delta$}}}{ \alpha + (1-\alpha)\underbrace{(1-F_{N}(0, \mu, \sigma)}_{\text{Truncated Normal Distribution over $[0,\infty)$}}}  \label{Eq: MixCDFInfinity} \\
		& \text{where $0 \leq \delta \leq \infty$} \nonumber
	\end{align}
	Since the debiased empirical delay distribution is only defined on $[0,\delta_{\max}]$, which is the domain on which it is compared to the modeled delay distribution, the truncation of the modeled distribution to this domain is also defined, denoted by $F'_\theta$ as below.
	\begin{align}
		F'_\theta(\delta)= &\frac{\alpha (F_{Exp}(\delta, Scale)) + (1-\alpha)\overbrace{(F_{N}(\delta, \mu, \sigma)-F_{N}(0, \mu, \sigma)}^{\text{Truncated Normal Distribution until $\delta$}}}{ \alpha (F_{Exp}(\delta_{\max}, Scale)) + (1-\alpha)\underbrace{(F_{N}(\delta_{\max}, \mu, \sigma)-F_{N}(0, \mu, \sigma)}_{\text{Truncated Normal Distribution over $[0,\delta_{\max}]$}}} \label{Eq: MixCDFDeltaMax}\\
		&\text{where $0 \leq \delta \leq \delta_{\max}$} \nonumber
	\end{align}
	
	\subsection{Defining the Optimization Function}\label{ssec:OptimizationFunction}
	Defining the optimization function was challenging due to two factors in particular:
	\begin{itemize}
		\item{There are many combinations of parameters that give approximately the same distribution when restricted to the domain $[0,\delta_{\max}]$, but which differ substantially in how much of the total distribution's weight is contained in this domain.\label{fctr: FirstFactor}}
		\item{As two year windows closer to the present are considered, the quantity of data shrinks, resulting in increasingly unstable parameter estimates.\label{fctr: SecondFactor}}
	\end{itemize}
	
	The optimization function used is shown in Eq.~\ref{eq:OptFunc} below. The first term\footnote{$F'_\theta$ computed as defined in Eq.~\ref{Eq: MixCDFDeltaMax}} $\|\log_{10}F'_\theta - \log_{10}F_\Delta  \|^2$  reduces the CDF difference between the debiased empirical delay distribution and the modeled delay distribution over the domain $[0,\delta_{\max}]$. The purpose of applying $\log_{10}$ weights is to place more emphasis on a good fit for the initial months\footnote{ The rationale behind $\log_{10}$, is to obtain CDF values close from the point of the ratio between the two distributions (modeled and debiased empirical), not in terms of absolute difference - a $\log_{10}$ CDF difference between 0.03 and 0.06 would be a factor of two difference in the correction whereas a difference between 0.93 and 0.96 would be much smaller despite the fact that the absolute difference is 0.03 in both cases.}. As mentioned above, there are many different combinations of parameters that would result in comparable errors in the first optimization term, but which nevertheless differ substantially over the domain $[0, \infty)$. This problem is relatively minor when the range $[0, \delta_{\max}]$ contains the bulk of the distribution, which it does when $\delta_{\max}$ is substantially greater than the second peak of the debiased empirical delay distribution. But as two year windows closer to the present are considered, this ceases to be the case. In order to avoid this problem, the optimization function must consider modeled delay distribution values beyond $\delta_{\max}$. The second term $\|\log_{10}S_\theta - \log_{10}S_{\theta'}\|^2$ penalizes large differences between the CDFs of consecutive modeled distributions beyond $\delta_{\max}$. In order to further minimize parameter instability for recent two year windows, another set of weights is assigned to the first two terms, which effectively diminishes the importance of a good fit with the empirical data and increases the importance of parameter stability as the amount of available data diminishes.
	
	The third and fourth terms are penalization terms - $F_N^2(0, \mu, \sigma)$ term penalizes negative delays introduced by the normal distribution (defined over $(-\infty, +\infty)$) whereas $S_\theta^2(10Y)$ term penalizes delays beyond 10 years.
	
	Mathematically, the optimization function is defined as
	\begin{equation}
		\begin{aligned}
			\theta_{Opt} & =
			\underset{\theta=(\alpha, Scale, \mu, \sigma)}{argmin}\frac{\delta_{\max}}{\delta_{Fix}}\underbrace{\|\log_{10}F'_\theta - \log_{10}F_\Delta  \|^2}_{\text{$ \delta \in [0,\delta_{\max}]$}} \\
			& \qquad \qquad \qquad + \left(1-\frac{\delta_{\max}}{\delta_{Fix}}\right) \underbrace{\|\log_{10}S_{\theta'}-\log_{10}S_\theta\|^2}_{\text{$\delta \in (\delta_{\max},  \delta_{Fix}]$}}\\
			& \qquad \qquad \qquad + \underbrace{F_N^2(0, \mu, \sigma)}_{\text{$\delta < 0$}} +  \underbrace{S_\theta^2(10Y)}_{\text{$\delta >10 Years$}}\\
			&\text{where $\delta_{Fix}$ is the Maximum value of $\delta$ in the dataset.}
		\end{aligned} \label{eq:OptFunc}
	\end{equation}
	
	$F'_\theta$ is as defined in Eq.~\ref{Eq: MixCDFDeltaMax} whereas $S_\theta$ is the survival function, defined as the complement of $F_\theta$ of Eq.~\ref{Eq: MixCDFInfinity}:
	\begin{align}
		S_{\theta} &= 1 - F_{\theta}
	\end{align} 
	
	Finally, $\theta'$ refers to the previous two year window's optimal parameters. Since no previous parameters are available for the first two year window, the second term is taken to be zero for that window.
	\begin{align}
		\|\log_{10}S_{\theta'}-\log_{10}S_\theta\|^2 = 0  \label{Eq:ZeroCDFDifference}  \\
		\text{$\theta'$ refers to optimal parameters at previous step.} \nonumber
	\end{align}
	
	The covariance matrix adaptation evolution strategy (CMA-ES) was applied to compute the modeled distribution parameters. It is a derivative free optimization algorithm, a type typically used when derivatives are difficult or costly to compute~(\cite{Hansen2006, Hansen2016, Hansen}).
	
	\subsection{Computing the Corrected Counts} \label{ssec:CorrectedCounts}
	Once the modeled delay distributions (one for each two year window) have been obtained by optimization, event counts can be corrected based on the cumulative distribution function of the full modeled distribution (i.e., defined over $[0, \infty)$) defined by Eq.~\ref{Eq: MixCDFInfinity}.

	Specifically, 
	\begin{align}
		\text{Corrected Count} &= \frac{\text{Reported Counts for month, `m'}}{F_\theta(a)} \label{Eq: CorrectedCounts}\\
		& \text{where $a$  is age of the given month, `m'.}\nonumber
	\end{align}
	
	From each modeled distribution, the correction factor is computed at only one point, $age$, of the modeled distribution for the given month's correction. 
	
	\section{Results}
	\label{sec:Result}
\begin{figure}[htbp]
		\centering
		\begin{subfigure}{.5\textwidth}
			\centering
			\fbox{\includegraphics[scale=0.3, angle=90,origin= c]{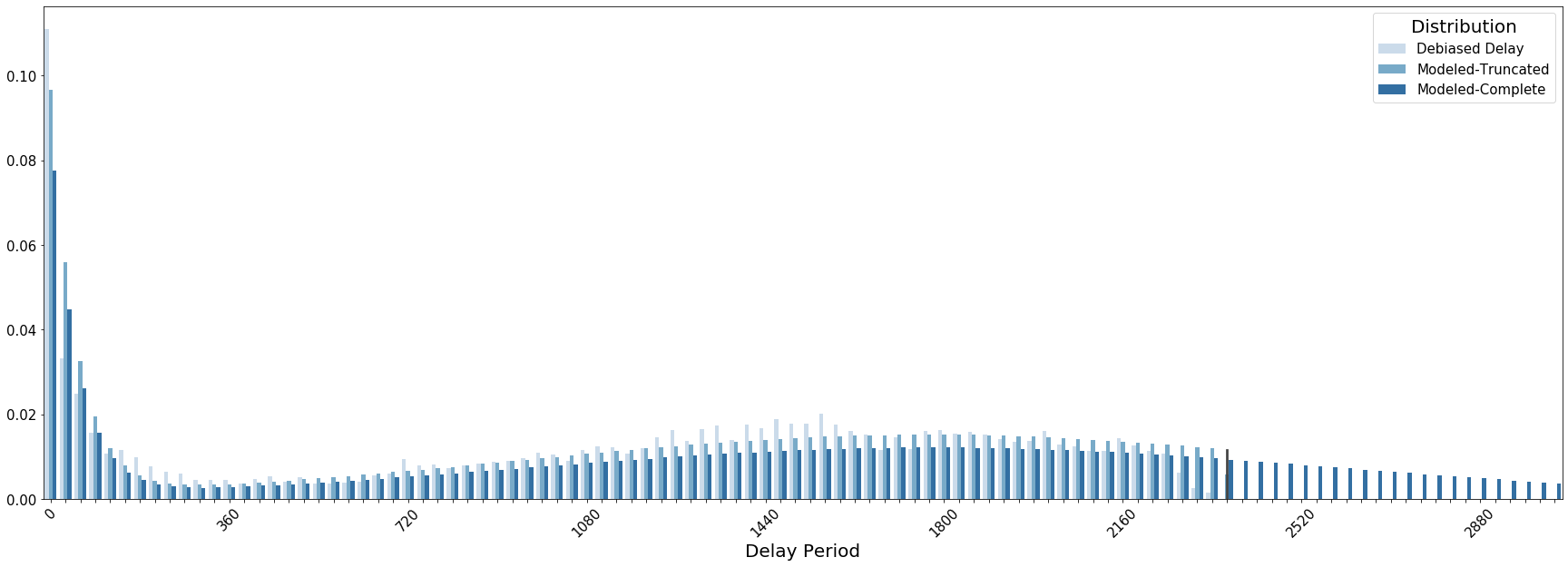}}
			\caption{From July 2012 to June, 2014}
			\label{fig:062014PDF}
		\end{subfigure}%
		\begin{subfigure}{.5\textwidth}
			\centering
			\fbox{\includegraphics[scale=0.3,  angle=90,origin= c]{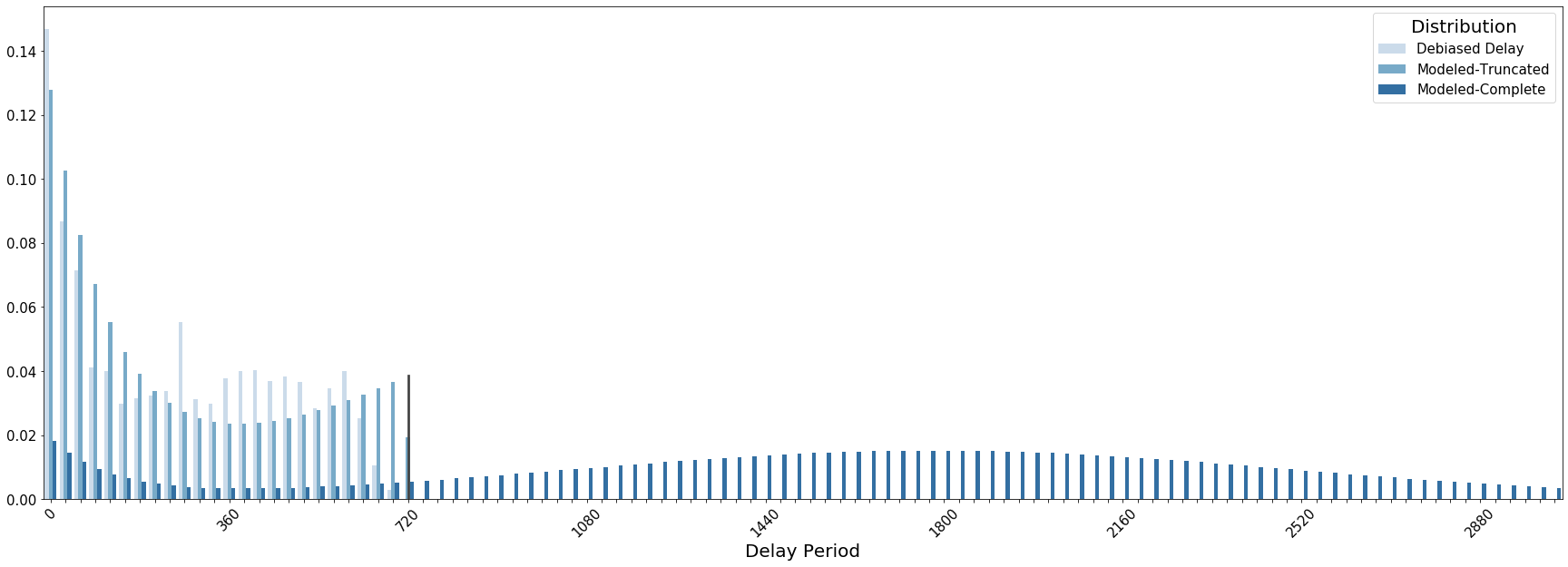}} 
			\caption{From January, 2017 to December, 2018}
			\label{fig:122018PDF}
		\end{subfigure}
		\vspace{-12pt}
		\caption{Comparing PDFs of Debiased Delay Distribution with Parametric Modeled Distribution}
		\label{fig:PDFMatching}
	\end{figure}

	Fig.~\ref{fig:PDFMatching} shows two examples of plots comparing PDFs of the fitted parametric modeled distribution, its truncation to the domain $[0, \delta_{\max}]$ --- where $\delta_{max}$ is computed for each window individually, and the debiased empirical delay distribution. Fig.~\ref{fig:062014PDF} shows this comparison for the two year window starting from July 2012 until June 2016 and Fig.~\ref{fig:122018PDF} shows this comparison for the most recent window starting from January, 2017 until December 2018.
	
	\subsection{Parameters and Interpretation} \label{ssec:Parameters}
\begin{figure}[ht]
		\centering
		\begin{subfigure}{.5\textwidth}
			\centering
			\fbox{\includegraphics[width=0.95\linewidth]{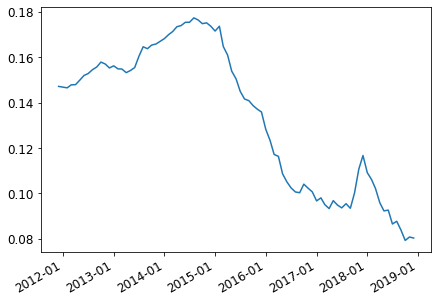}}
			\caption{Alpha Parameter}
			\label{fig:AlphaPlot}
		\end{subfigure}%
		\begin{subfigure}{.5\textwidth}
			\centering
			\fbox{\includegraphics[width=0.95\linewidth]{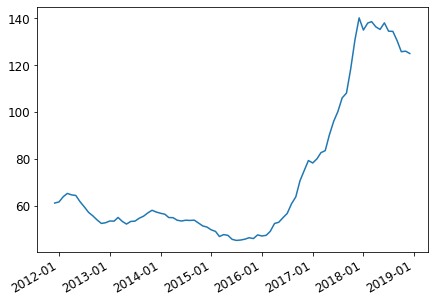}}
			\caption{Scale Parameter}
			\label{fig:ScalePlot}
		\end{subfigure} 
		\begin{subfigure}{.5\textwidth}
			\centering
			\fbox{\includegraphics[width=0.95\linewidth]{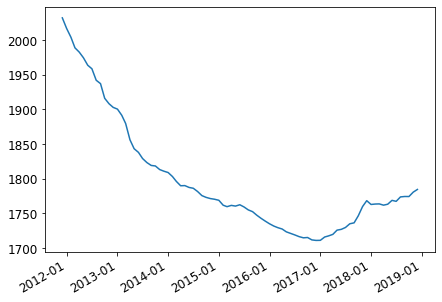}}
			\caption{Mean Parameter}
			\label{fig:MeanPlot}
		\end{subfigure}%
		\begin{subfigure}{.5\textwidth}
			\centering
			\fbox{\includegraphics[width=0.95\linewidth]{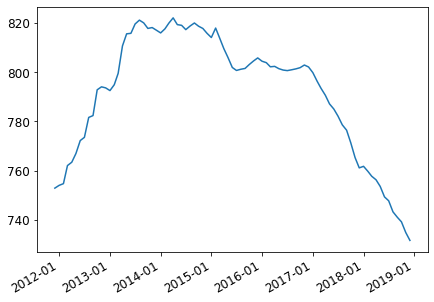}}
			\caption{Sigma Parameter}
			\label{fig:SigmaPlot}
		\end{subfigure}
		\vspace{-10pt}
		\caption{Plots of Modeled Distribution Parameters based on Empirical Debiased Delay Distribution}
		\label{fig:ParamterPlots}
	\end{figure}

	Fig.~\ref{fig:ParamterPlots} shows the parameter plots of the delay distribution generated for each monthly two year rolling window.
	
	The alpha plot (Fig.~\ref{fig:AlphaPlot}) suggests that organizations discover 8-18\% of cyber events right away ($8\% \leq \alpha \leq 18\%$).
	
	The scale plot (Fig.~\ref{fig:ScalePlot}) suggests that the short delays modeled by the exponential distribution had a mean of less than 60 days delay until early 2016 but increased rapidly to around 140 days in early 2018.
	
	The normal distribution mean, $\mu$, (Fig.~\ref{fig:MeanPlot}) and standard deviation, $\sigma$, (Fig.~\ref{fig:SigmaPlot}) parameter plots suggest that the longer delays modeled by the normal distribution remained consistent over time. The period of longer delays remain consistent varying within $\pm 10\%$ range.
	
	\subsection{Corrections and Validation} \label{ssec: Validation}
\begin{figure}[htbp]
		\centering
		\begin{subfigure}{.5\textwidth}
			\centering
			\fbox{\includegraphics[scale=0.292, angle=90,origin= c]{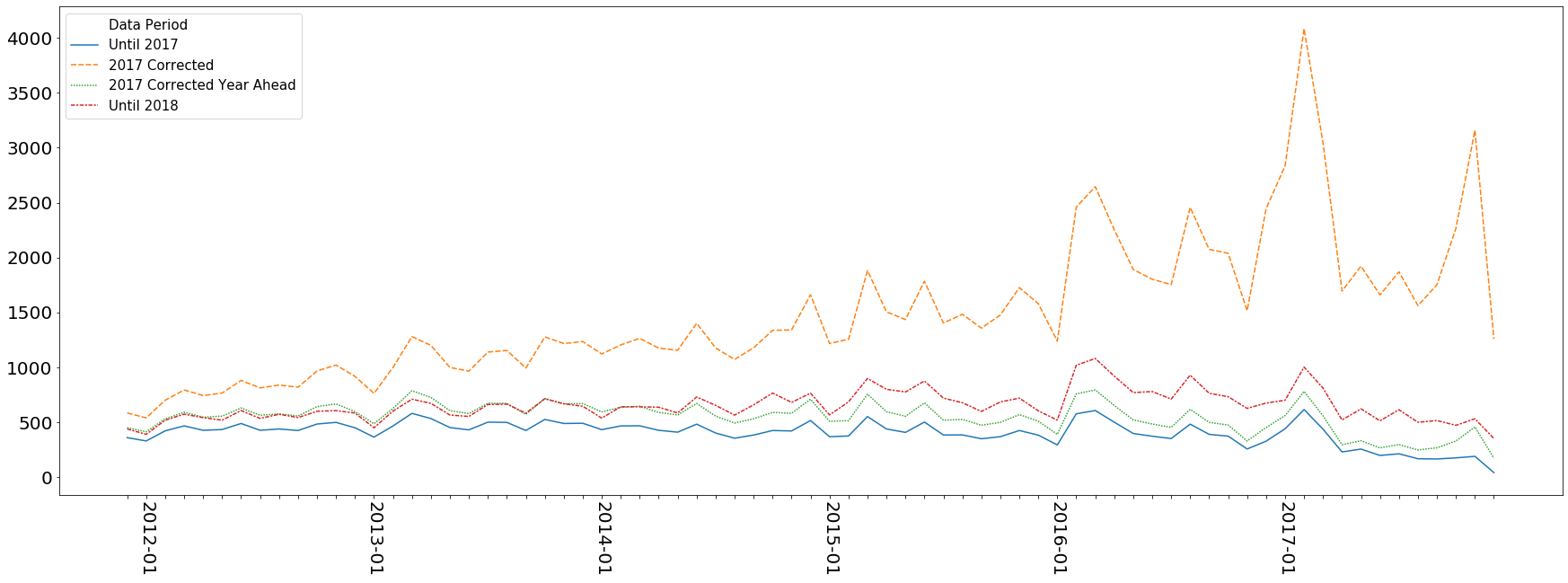}}
			\caption{2017 corrections Vs 2018 cumulative counts}
			\label{fig:Validation2017}
		\end{subfigure}%
		\begin{subfigure}{.5\textwidth}
			\centering
			\fbox{\includegraphics[scale=0.292,  angle=90,origin= c]{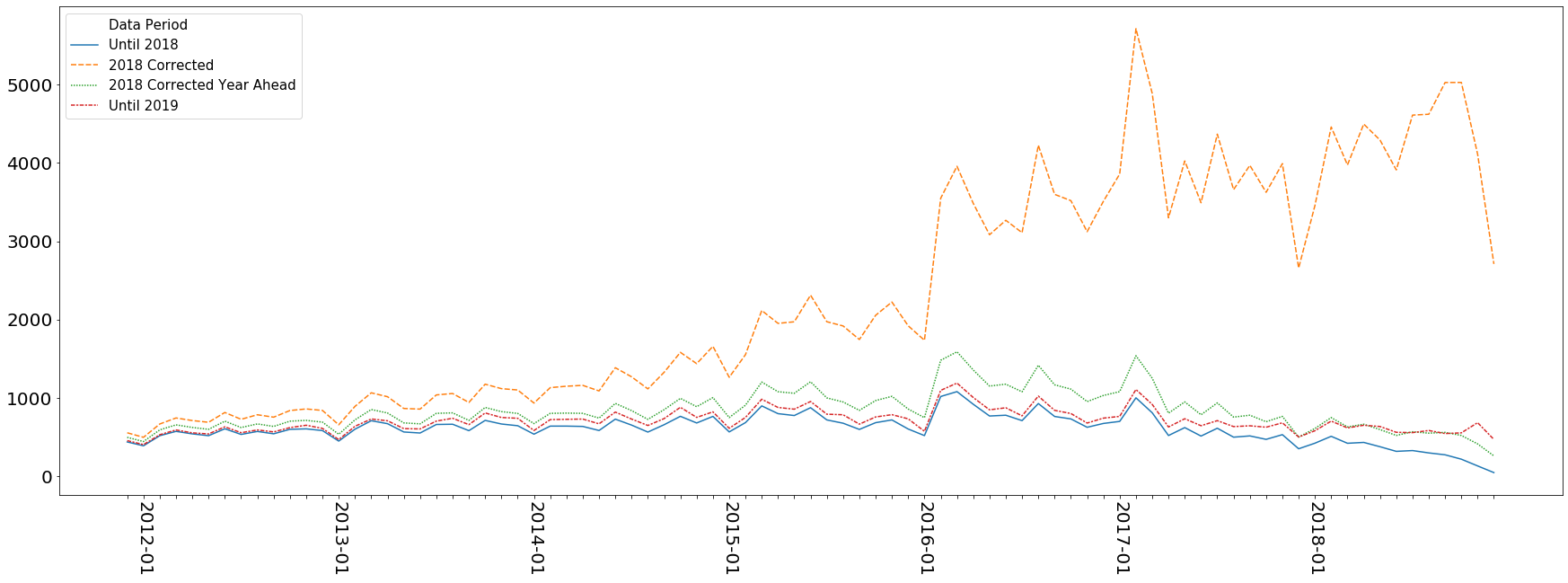}} 
			\caption{2018 corrections Vs 2019 cumulative counts}
			\label{fig:Validation2018}
		\end{subfigure}
		
		\begin{align*}
			& \text{{ \footnotesize
					Until 201X - Counts reported as of 201X adjusted for the default date of January 1, proportionally}} \\[-0.8 em]
			& \text{{ \footnotesize
					201X Corrected - ``Counts until 201X" corrected based on Eq.~\ref{Eq: CorrectedCounts}}} \\[-0.8em]
			& \text{{ \footnotesize
					201X Corrected Year Ahead - ``Counts until 201X" corrected based on Eq.~\ref{eq:CorrectedProbYearAhead}}}
		\end{align*}
		\vspace{-15pt}
		\caption{Validation Plots}
		\abovedisplayskip=-15pt
		\label{fig: Validations}
	\end{figure}

	Fig.~\ref{fig: Validations} shows the corrected incident counts based on the proposed methodology.
	Figs.~\ref{fig:Validation2017} and~\ref{fig:Validation2018} show the corrected counts for the events reported by Dec. 2017 and by Dec. 2018, respectively. Although the corrections follow similar trends in both, the correction factors vary substantially.
	
	To validate the proposed algorithm, the counts reported until December 2017 (2018) were corrected for a year ahead and compared against the counts reported as of December 2018 (2019).
	The year ahead correction factor is computed as 
	\begin{equation}
		\text{ Year ahead }F_\theta(a, a+1 \: Year) =\frac{F_\theta(a)}{	F_\theta(a+1 \: Year)} \label{eq:CorrectedProbYearAhead}
	\end{equation}
	\qquad where $a$ is the age of the event counts being corrected.
	
	Whereas the 2017 year ahead corrections (Fig.~\ref{fig:Validation2017}) initially show close agreement with the 2018 counts, more recent year ahead corrections underestimate the 2018 counts. On the other hand, the 2018 year ahead corrections (Fig.~\ref{fig:Validation2018}) generally overestimate the 2019 counts, except for the most recent months, which show close agreement. As stated in section~\ref{ssec:OptimizationFunction}, the debiased empirical delay distribution has fewer data points for more recent two year windows so weights of ($\sfrac{\delta_{\max}}{\delta_{Fix}})$ and ($1 - \sfrac{\delta_{\max}}{\delta_{Fix}}$) are used in the optimization function to dynamically adjust the weight given to the CDF before and after $\delta_{\max}$ respectively. By removing these weights, better estimates for recent months might be obtained but would come at the cost of more parameter instability and worse validation plots (overfitting). 
	In either Fig.~\ref{fig:Validation2017} or Fig.~\ref{fig:Validation2018}, the corrected counts (dashed line) show a trend of increasing incident counts since 2016, which is contrary to the diminishing trend seen in the raw counts. The trend in corrected counts is therefore much more in line with reports from insurers and other organizations that release reports on cyber risk.
	
	\section{Conclusion}
	\label{sec:conc}
	This work examined the long known problem of reporting delays in historical cyber events databases and proposed an algorithm to correct for these delays. Interestingly, the true distribution of reporting delays appears to be bi-modal, which we have interpreted as a mixture of two distributions: one for incidents that are discovered immediately, modeled by an exponential distribution, and one for incidents that are not immediately discovered, modeled by a normal distribution. With this form of reporting delay distribution, we obtained non-stationary modeled delay distributions via optimization. These modeled delay distributions were used to estimate the total number of cyber incidents that will eventually be reported from the current counts. The approach was validated by estimating year ahead corrections.
	
	To understand the current cyber threat landscape and to create robust cyber risk models, one needs accurate historical data. While it is not possible to get the exact count of cyber events, the proposed algorithm aims to correct for reporting delays approximately. The reported cyber incident counts in recent times show a decreasing trend simply because incidents have not been reported yet, even though they have actually already occurred. However, in reality, the rate of cyber incidents is increasing and that is what the algorithm reveals.
	
	The general approach can be applied to any form of data with reporting delays including other long tailed insurance claims (such as liability), and COVID-19 pandemic data. The current study corrects the overall counts for US cyber events, and the industry specific correction of cyber event counts is a future direction of this work.
	
	\section*{Acknowledgment}
	The study was done in collaboration with AIR Worldwide using their proprietary cyber data. The authors would like to thank Scott Stransky for his comments and suggestions that helped improve the approach.
		\bibliographystyle{plainnat}

\end{document}